\RequirePackage{fix-cm}
\RequirePackage{amsmath}
\documentclass[twocolumn,epjc3]{svjour3}  

\smartqed  % flush right qed marks, e.g. at end of proof
\RequirePackage{graphicx}
\RequirePackage{mathptmx}      % use Times fonts if available on your TeX 
\usepackage{color}
\usepackage{mathrsfs}
\usepackage{amssymb, bm}

\newcommand{\be}{\begin{equation}}
\newcommand{\ee}{\end{equation}}

\begin{document}

\title{Cosmological analogies for geophysical flows, Lagrangians, and new 
analogue gravity systems
  %\thanksref{t1}
}
%\subtitle{}


\author{Valerio Faraoni\thanksref{e1,addr1}
	   \and
Sonia Jose \thanksref{e2,addr1} 
}

\thankstext{e1}{e-mail: vfaraoni@ubishops.ca}
\thankstext{e2}{e-mail: sjose21@ubishops.ca}


\institute{Department of Physics \& Astronomy, Bishop's University, 
2600 College Street, Sherbrooke, Qu\'ebec, Canada J1M~1Z7 \label{addr1}
	%\emph{Present Address:} if needed\label{addr3}
}

\date{Received: date / Accepted: date}

\maketitle

\begin{abstract}

Formal analogies between the ordinary differential equations describing 
geophysical flows and Friedmann cosmology are developed. As a result, one 
obtains Lagrangian and Hamiltonian formulations of these equations, while 
laboratory experiments aimed at testing geophysical flows are shown to 
constitute analogue gravity systems for cosmology.

\end{abstract}

\section{Introduction}
\label{sec:1}
\setcounter{equation}{0}

There are unexpected formal analogies between cosmology and geophysical 
flows. Geophysical flows encompass many natural phenomena, including lava 
flows, 
the creep of glacier ice, avalanches, and mud slides. In particular, the 
analogy between early models of ice caps on a horizontal bed (based on an 
incorrect ice rheology) \cite{Nye51} and lava domes is well known 
\cite{Blake90,Griffithsreview}. These early models of ice caps treated 
glacier ice 
as a perfectly plastic material, {\em i.e.}, a non-Newtonian fluid that 
does not yield under stress until a certain threshold (``yield stress'') 
is reached, at which point the material deforms abruptly.  This kind of 
material is nowadays referred to as a Bingham fluid. It is now established 
that glacier ice instead deforms under stress according to the non-linear 
Glen law relating the strain rate tensor $\dot{\epsilon}_{ij}$ and the 
deviatoric stresses $ \hat{s}= ( s_{ij} ) $ \cite{Glen} 
\be 
\dot{\epsilon}_{ij} = {\cal A} \, \sigma_\mathrm{eff}^{n-1} \, s_{ij} 
\,,\label{Glen} 
\ee 
where ${\cal A}$ is a constant (that depends on the 
temperature, crystal orientation, and impurities 
\cite{Paterson,CuffeyPaterson,Hooke,GreveBlatter})  and 
\be 
\sigma_\mathrm{eff} = 
\sqrt{ \frac{1}{2} \, \mbox{Tr} \left( \hat{s}^2 \right)} 
\ee 
is the effective 
stress \cite{Paterson,CuffeyPaterson,Hooke,GreveBlatter}. However, Nye's discussion 
applies without change to a Bingham fluid on a horizontal bed, which is a 
good model for lava flow, and gives the parabolic profile for a lava dome 
on horizontal bed \cite{Blake90,OsmondGriffiths98,Griffithsreview}. (The Bingham
 fluid is the most 
common non-Newtonian fluid model to describe lava flows 
\cite{Robson67,Johnson70,Hulme74,PinkertonSparks78,Dragonietal86,KilburnLuongo93,TallaricoDragoni99,TallaricoDragoni00}.) Similarly, the discussion of 
perfectly plastic glacier ice on a slope, although inadequate to describe 
an alpine glacier because of the wrong rheology, describes a lava flow on 
a slope. The corresponding analytical solution of the relevant 
fluid-mechanical equations appears in the pedagogical literature as a 
simple example of how different ice rheologies produce different 
macroscopic glacier profiles \cite{EJP}. Unbeknownst to the author of 
\cite{EJP}, this solution is perfectly adequate to describe the flow of 
lava (a Bingham fluid) on a slope \cite{Griffithsreview}. Below, we 
elaborate on this analogy. 

The ordinary differential equations ruling the longitudinal profiles of 
glaciers, ice caps, lava flows, or lava domes lend themselves to  
analogies with the Einstein-Friedmann equations of cosmology. These 
equations describe the evolution of a spatially homogeneous and isotropic 
universe in general relativity and constitute the basis of modern (or 
Friedmann-Lema\^itre-Robertson-Walker, in short ``FLRW'') cosmology. The 
various solutions describing geophysical flows correspond to different 
matter contents and curvatures for these universes. Since Lagrangian and 
Hamiltonians for the equations of FLRW cosmology are known, the cosmic 
analogy provides a way to identify Lagrangians and Hamiltonians for the 
differential equations describing the analogous geophysical flows. While 
analogies between geophysical (and other) systems have been explored in 
the literature (see \cite{mybook} for a review), the ones that we examine 
here are novel.

In the next section we recall the basics of FLRW cosmology for the reader 
unfamiliar with it. Sec.~\ref{sec:3} discusses a Newtonian fluid model of 
a lava front and the relevant cosmic analogy, Lagrangian, and Hamiltonian. 
Section~\ref{sec:4} studies a Bingham fluid model of a lava front; 
Sec.~\ref{sec:5} discusses Bingham models of lava domes and ice caps on 
horizontal beds and their cosmic analogues, while Sec.~\ref{sec:6} extends 
the discussion to flows on a slope and Sec.~\ref{sec:7} contains the 
conclusions.

\section{Basics of FLRW cosmology}
\label{sec:2}

We follow the notation of  Refs.~\cite{Wald,Carroll} and  use units in 
which the speed  of light is unity. $G$ 
is Newton's constant and the four-dimensional metric tensor has signature ${-}{+}{+}{+}$.

Under the strong mathematical requirements of spatial homogeneity and 
isotropy motivated by observations of the cosmic microwave background 
permeating our universe and by large-scale structure surveys, the 
four-dimensional geometry of the universe is necessarily given by the FLRW 
line element, which reads \cite{MTW,Wald,Carroll,KT,Liddle}
\begin{eqnarray}
ds^2 &=& g_{\mu\nu} \, dx^{\mu} dx^{\nu} \nonumber\\
&&\nonumber\\
&=& -dt^2 +a^2(t) \left[ \frac{dr^2}{1-Kr^2} +r^2 \left( d\vartheta^2 + 
\sin^2 \vartheta \, d\varphi^2 \right)\right] \nonumber\\
&& \label{eq:10}
\end{eqnarray}
in comoving polar coordinates $x^{\mu}= \left(t, r, \vartheta, \varphi  
\right)$, 
where $g_{\mu\nu}$ is the metric tensor. 
The  scale factor $a(t)$ describes the expansion history of 
the universe, while the constant $K$  describes the  
curvature of the 3-dimensional spatial sections (the 3-geometries obtained 
by setting  $dt=0$). If $K>0$ the line 
element~(\ref{eq:10}) describes closed universes; if $K=0$ it corresponds 
to Euclidean (flat) spatial 
sections and, if $K<0$, it describes  hyperbolic 3-spaces 
\cite{Wald,Carroll,Liddle,KT}. All the 
possible FLRW geometries fall in these three categories classified by the 
sign of the curvature index $K$. 

The cosmological spacetime is curved by its mass-energy content and 
different matter contents produce different cosmic histories $a(t)$.  
The cosmic matter is usually described by a perfect fluid with energy 
density $\rho(t)$ and isotropic pressure $P(t)$ related by a barotropic  
equation  of state $P=P(\rho)$. The scale factor $a(t)$ 
and the matter variables $\rho(t), P(t)$ satisfy the Einstein-Friedmann 
equations ({\em i.e.}, the Einstein equation of general relativity adapted 
to the symmetric line element~(\ref{eq:10}) 
\cite{Wald,Carroll,Liddle,KT}  
\begin{eqnarray}
&&H^2 \equiv \left( \frac{\dot{a}}{a}\right)^2 =\frac{8\pi G}{3} \, \rho 
-\frac{K}{a^2} \,, \label{eq:11}\\
&&\nonumber\\
&&\frac{\ddot{a}}{a}= -\, \frac{4\pi G}{3} \left( \rho +3P \right) \,, 
\label{eq:12} \\
&&\nonumber\\
&& \dot{\rho}+3H\left(P+\rho \right)=0 \,,\label{eq:13}
\end{eqnarray}
where an overdot denotes differentiation with respect to the comoving 
time $t$ while $H(t)\equiv 
\dot{a}/a$ is the Hubble function 
\cite{Wald,Carroll,Liddle,KT}.
Given any two of these equations, the third one can be 
derived from them so that only 
two equations are independent. For convenience, here we choose the 
Friedmann equation~(\ref{eq:11}) and the energy conservation 
equation~(\ref{eq:13}) as primary, regarding the acceleration 
equation~(\ref{eq:12}) as derived.  Therefore, our analogies between 
geophysical flows and FLRW cosmology will be valid only if   the energy 
conservation equation~(\ref{eq:13}) is satisfied in addition to the 
Friedmann equation~(\ref{eq:11}) 
(which is easy to verify). This happens when a  cosmological fluid 
satisfying the covariant conservation equation~(\ref{eq:13}) fills the 
analogous universe. If this fluid has barotropic equation of state 
$P=w\rho$  with $w=$~const., the conservation equation~(\ref{eq:13}) corresponds to 
an energy density scaling as 
\be
\rho(t)=\frac{\rho^{(0)}}{ \left[ a(t) \right]^{3(w+1)}  } 
\,,\label{eq:14} 
\ee 
where $\rho^{(0)} $ is a positive integration constant determined by the 
initial conditions \cite{MTW,Wald,Carroll,Liddle,KT}. Therefore, to 
establish the validity of an analogy, 
it is sufficient to establish the validity of an equation of the form of  
the Friedmann equation with a fluid source satisfying Eq.~(\ref{eq:14}), 
or sourced by a mixture of (mutually decoupled) fluids each satisfying Eq.~(\ref{eq:14}).

The Lagrangian and Hamiltonian reproducing the Einstein-Friedmann equation 
through the Euler-Lagrange or the Hamilton equations are obtained from the 
action of general relativity with a perfect fluid \cite{Wald,Carroll}
\begin{eqnarray}
S &=&\int d^4 x \, \sqrt{-g^{(4)}} \left( \frac{R}{16\pi G} +\rho  \right) 
\nonumber\\
&&\nonumber\\
&=& 4\pi \int dr\, \frac{r^2}{\sqrt{1-Kr^2}}   \,\int dt \,  L\left( a, 
\dot{a} 
\right) 
\end{eqnarray}
(where $g^{(4)}$ is the determinant of the metric $g_{\mu\nu}$), which 
yields 
\begin{eqnarray}
L \left(a, \dot{a} \right) &=& \frac{3}{8\pi G} \left( a \dot{a}^2 -Ka 
\right) +a^3 \rho \,,\\
&&\nonumber\\
{\cal H} \left(a, \dot{a} \right) &=& \frac{3}{8\pi G} \left( a \dot{a}^2 
+ Ka \right) -a^3 \rho \,,
\end{eqnarray}
where the dynamics is constrained and the ``scalar'' or 
``Hamiltonian'' constraint ${\cal 
H}=0$ must be satisfied \cite{MTW,Wald,Carroll}. We are now  ready to 
build 
analogies between geophysical flows and FLRW 
cosmology.

\section{Newtonian model of a lava front}
\label{sec:3}

Lava behaves as a Newtonian fluid near a vent, where it is hotter, but 
sometimes also the front of a lava flow is modelled as a Newtonian fluid 
\cite{Griffithsreview}. An analytical Newtonian lava flow model is given 
in \cite{DragoniBorsariTallarico05} and its analogy with FLRW cosmology 
was mentioned in Ref.~\cite{mybook}, which we report and complete here.

Assume that the lava front is homogeneous and isothermal, that it  
moves at constant velocity on an inclined plane, and that it extends 
indefinitely in the transversal ($y$-) direction. Let $x$ be a coordinate 
down-slope, $h(x)$ be the lava thickness measured along an axis 
perpendicular to the bed, and $L$ be the length of the flow, and assume 
$h\ll L$ (this shallow fluid approximation is common in glaciology and in 
the study of geophysical flows).  Let $\rho$ and $\eta $ be the lava 
density and dynamic viscosity coefficient, $g$ the acceleration of 
gravity, $\beta $ the slope of the plane lava bed, and $v_0$ the 
(constant) velocity of the lava in a reference frame fixed to the ground 
\cite{DragoniBorsariTallarico05}. When lava is described a Newtonian 
fluid, the Navier-Stokes equations for laminar flow provide the 
differential equation for the lava flow profile $h(x)$ 
\cite{DragoniBorsariTallarico05}
\be
b_0 \, h^3 h' -a_0 \, h^3 +v_0 \, h =0 \label{8eq:lavaflow}
\ee
where $ h' \equiv dh/dx $ and 
\begin{eqnarray}
a_0 &=& \frac{ \rho g \sin\beta}{3\eta} \,,\\
&&\nonumber\\
b_0 &=& \frac{ \rho g \cos\beta}{3\eta} \,.
\end{eqnarray}
The analogy with FLRW cosmology follows from rewriting this equation as  
\be
\left( \frac{h'}{h} \right)^2  = \frac{a_0^2}{b_0^2 \, h^2} + 
\frac{v_0^2}{b_0^2 \, h^6} -\frac{2a_0 v_0}{b_0^2 \, h^4} 
\,,\label{8eq:azzoazzo}
\ee
which is analogous to the Friedmann equation
\be
H^2 = -\frac{K}{a^2} +\frac{8\pi G \rho_\mathrm{(stiff)} }{3a^6} + 
\frac{8\pi G \rho_\mathrm{(rad)}  }{3a^4} 
\ee
for a universe with negative curvature filled by a stiff fluid with 
equation of state $P_\mathrm{(stiff)}=\rho_\mathrm{(stiff)} $ and a 
radiation fluid with $P_\mathrm{(rad)}=\rho_\mathrm{(rad)} /3$ (which has 
negative energy 
density, a fact that would be unacceptable in realistic cosmology but is 
rather immaterial in our formal analogy). The map between lava front and 
cosmology reads  
\begin{eqnarray}
&& K = - \left( \frac{a_0}{b_0} \right)^2 = - \tan^2 \beta \,,\\
&&\nonumber\\
&& 8\pi G \rho_\mathrm{(stiff)} = \frac{3 v_0^2}{ b_0^2}
= 3\left( \frac{3\eta v_0}{\rho g \cos\beta}\right)^2  \,,\\
&&\nonumber\\
&& 8\pi G \rho_\mathrm{(rad)} = -\frac{ 6a_0 v_0}{ b_0^2}
=- \frac{ 18 v_0 \eta \sin\beta}{\rho g \cos^2\beta}  <0 \,.  
\end{eqnarray}
The well-known Lagrangian of the analogous FLRW universe indicates 
the Lagrangian for the lava flow problem 
\be
L \left( h,h'\right) = h \, h'^2 -\frac{2a_0 v_0}{b_0^2 h} 
+\frac{v_0^2}{b_0^2 h^3} +\left( \frac{a_0}{b_0} \right)^2 h \,.
\ee
Since $ L $ does not depend explicitly on $x$, the corresponding 
Hamiltonian
\be
{\cal H}= h \, h'^2 + \frac{2 a_0 \, v_0}{ b_0^2 h} 
-\frac{v_0^2}{ b_0^2 h^3} - \left( \frac{ a_0}{ b_0} \right)^2 h
\ee
is conserved. Equation~(\ref{8eq:azzoazzo}) for the lava flow profile is 
recovered by setting $ {\cal H}=0 $ (this is the Hamiltonian constraint 
of the Einstein equations).

An analytic solution of Eq.~(\ref{8eq:lavaflow}) is  
\cite{DragoniBorsariTallarico05} 
\be
x_0 - x = H_0 \cot \beta \left[ \tanh^{-1} \left( \frac{h}{H_0} 
\right) -\frac{h}{H_0} \right] \,,\label{eq:3.11} 
\ee
where  $ 0 \leq x \leq x_0 $ and 
\be
H_0 = \sqrt{ \frac{3\eta \, v_0}{\rho g \sin\beta}}  
= \sqrt{ \frac{v_0}{a_0} } 
= \sqrt{ \frac{ 2\rho_\mathrm{(stiff)} }{ |\rho_\mathrm{(rad)}|}  } \,.
\ee
This equation provides the scale factor $a(t)$ of the spatially curved 
analogous universe
\be
\frac{t_0 - t}{\tau} = \tanh^{-1} \left( \frac{a}{a_*} \right)  
-\frac{a}{a_*}  
\ee
for $ 0\leq t \leq t_0 $, where 
\be
\tau = \sqrt{ \frac{ 2\rho_\mathrm{(stiff)} }{ |\rho_\mathrm{(rad)} |} }\, 
\cot\beta = \sqrt{ \frac{2\rho_\mathrm{(stiff)} }{\left| 
\rho_\mathrm{(rad)} K \right|} } 
\ee
and $a_* = \sqrt{  2\rho_\mathrm{(stiff)} / \left| \rho_\mathrm{(rad)} 
\right| }$. 
In the limit  $v_0\to 0$, the solution for 
the (now solidified) lava front 
degenerates into 
the trivial straight line $h(x)= \left( x-x_0 \right) \tan \beta$. The 
analogous FLRW universe is empty and has hyperbolic three-dimensional 
spatial sections,  according to 
\be
H^2=-\frac{K}{a^2} \,,
\ee
and linear scale factor $a(t)=\sqrt{|K|} \, t$. This is empty Minkowski 
spacetime in a hyperbolic foliation ({\em i.e.}, in accelerated 
coordinates) in which the three-dimensional space is curved, while the 
four-dimensional curvature is identically zero \cite{Mukhanov,Liddle,KT}.

The other limit of the solution~(\ref{eq:3.11}) for $\beta\to 0$, in which 
the bed becomes horizontal, corresponds to zero spatial curvature and the 
stiff fluid as the only matter source in the cosmic analogy. This limit is 
interesting because it reproduces the shape of an accretionary wedge in 
the oceanic crust \cite{EmermanTurcotte83}. Setting $\beta =0$ gives 
\cite{DragoniBorsariTallarico05} the profile
\be
h(x) \simeq \left[ \frac{9\eta v_0}{\rho g} \left(x_0-x \right) 
\right]^{1/3} 
\ee
for $ 0\leq x \leq x_0 $. The scale factor of the analogous spatially flat 
expanding universe reduces to the well-known power-law $ a(t) \simeq a_* 
\left( t-t_0 \right)^{1/3} $ (with $a_*$ a positive constant) caused by a stiff fluid or  a free scalar 
field \cite{Faraoni:2021opj}.

\section{Bingham fluid model of a lava front}
\label{sec:4}

Away from a vent, where lava is cooler and more viscous, it behaves more 
like a Bingham fluid.  Consider now a Bingham model of  a lava front 
flowing down an incline with constant slope $\beta$. Let $\rho, \eta, 
\sigma_0, g $, and $v_0$ be the lava 
density, viscosity coefficient, yield stress, the acceleration of gravity, 
and the speed of the front, respectively. Then the Navier-Stokes equations give 
\cite{DragoniBorsariTallarico05}
\be
h'= \left( 1- \frac{3H_p}{2h} -\frac{H_N^2}{h^2} \right)  \tan \beta \,,
\ee
where
\be
H_p = \frac{\sigma_0}{\rho g \sin\beta} \,, \quad \quad 
H_N = \sqrt{ \frac{ 3\eta v_0}{\rho g \sin\beta} } \,.
\ee
Rearranging this equation one obtains 
\begin{eqnarray}
\left( \frac{h'}{h} \right)^2 &=& 
\frac{\tan^2\beta}{h^2} 
+ \left( \frac{9H_p^2}{4} - 2H_N^2 \right) \frac{\tan^2\beta}{h^4} 
+\frac{H_N^4 \tan^2\beta}{h^6}  \nonumber\\
&&\nonumber\\
&\, &  +\frac{3H_p H_N^2\tan^2\beta}{h^5} 
-\frac{3H_p\tan^2\beta}{h^3} \,.\label{analogBinghamfront}
\end{eqnarray}
In the analogous FLRW cosmos, the various terms in the right hand side of 
Eq.~(\ref{analogBinghamfront}) describe, respectively, hyperbolic 
curvature, radiation with energy density
$\rho_\mathrm{(rad)} =\rho_\mathrm{(rad)}^{(0)}/a^4$, a stiff fluid with 
$\rho_\mathrm{(stiff)} =\rho_\mathrm{(stiff)}^{(0)}/a^6$, a fluid with 
$w=2/3$, and a dust with zero pressure and 
$\rho_\mathrm{(dust)}=\rho^{(0)}_\mathrm{(dust)}/a^3$, where 
\begin{eqnarray}
K &=& -\tan^2\beta<0 \,,\\
&&\nonumber\\
\frac{8\pi G}{3} \,  \rho_\mathrm{(rad)}^{(0)} &=& \left( \frac{9H_p^2}{4} 
- 2H_N^2 \right) \tan^2\beta \nonumber\\
&&\nonumber\\
&= & \frac{3\left( 3\sigma_0^2-8\eta v_0 \rho g \sin\beta  \right)}{ 4 
\rho^2 g^2 
\cos^2\beta } \,,\nonumber\\
&&\\
\frac{8\pi G}{3} \,  \rho_\mathrm{(stiff)}^{(0)} &=& H_N^4 \tan^2\beta 
=\left( \frac{3\eta v_0}{\rho g \cos\beta} \right)^2 \,,\\
&&\nonumber\\
\frac{8\pi G}{3} \,  \rho_{(2/3)}^{(0)} &=&  3H_p H_N \tan^2\beta 
= \frac{3\sigma_0}{\cos^2\beta} \, \sqrt{ \frac{3\eta v_0 
\sin\beta}{\left( 
\rho g \right)^3} } \,,\nonumber\\
&&\\
\frac{8\pi G}{3} \,  \rho_\mathrm{(dust)}^{(0)} &=& -  3H_p \tan^2\beta 
=- \frac{3\sigma_0 \sin\beta}{\rho g \cos^2\beta} <0 \,.
\end{eqnarray}
Using the common values for lava $\sigma_0 \simeq 2000$~Pa, $\eta \simeq 
10^6 $~Pa$\cdot$s, $v_0\simeq 10^{-2}$~m$/$s, one obtains
$3\sigma_0^2 -4\eta v_0 \simeq 1.2 \cdot 10^7 $~Pa, making 
$\rho_\mathrm{(rad)}^{(0)} 
$ positive. However the dust fluid always has negative energy density.\\

\section{Bingham fluid models of lava domes and ice caps on horizontal 
beds}
\label{sec:5}

A Bingham fluid on a horizontal bed assumes a well-known parabolic profile 
found by Nye in early studies of ice caps \cite{Nye51}. Nye used 
the incorrect Bingham (or ``perfectly plastic'') rheology for ice, which 
was 
later superseded by Glen's law~(\ref{Glen}) \cite{Glen}, however the 
discussion applies 
without modification to Bingham fluids spreading on a horizontal 
background, such as a lava dome, and is confirmed by experiments 
\cite{Blake90,Griffithsreview}.

Consider an axisymmetric flow and let $x$ point in the radial direction, 
while 
$h(x)$ is the thickness of the (incompressible) material of density 
$\rho$ at $x$. The simplified Navier-Stokes equations for stationary 
state in the thin lava approximation give
\be
\frac{\partial P}{\partial x} = \rho g \, \frac{\partial h}{\partial x} 
\,,
\ee
where $g$ is the acceleration of gravity. The basal stress $ \tau_b =-\rho 
g  dh/dx $ is equated to the yield stress $\sigma_0$ everywhere  
\cite{Paterson,CuffeyPaterson,Hooke}, giving
\be
\rho g \, \frac{\partial h}{\partial x} = \frac{\sigma_0}{h} \,,
\label{quella}
\ee
which has as a solution the parabolic Nye profile 
\cite{Nye51,Paterson,CuffeyPaterson,Hooke} 
\be
h(x)=H\sqrt{ 1-\frac{x}{L} } \,, \quad \quad H= \sqrt{ 
\frac{2\sigma_0 \, L}{\rho g} }  \,.\label{Nyeprofile} 
\ee
By squaring, Eq.~(\ref{quella}) is written as 
\be
\left( \frac{h'}{h} \right)^2 = \left( \frac{\sigma_0}{\rho g} \right)^2 
\, \frac{1}{h^4} \,, \label{questa}
\ee
which is analogous to the Friedmann equation~(\ref{eq:11}) for a spatially 
flat ($K=0$) FLRW universe filled with blackbody radiation with equation 
of  state $P_\mathrm{(rad)}=\rho_\mathrm{(rad)} /3$ and energy 
density $\rho_\mathrm{(rad)}(t)=\rho^{(0)}_\mathrm{(rad)} /a^4$, in the correspondence
\be
h(x) \longleftrightarrow a(t) \,, \quad \quad 
\frac{8\pi G}{3} \, \rho^{(0)}_\mathrm{(rad)}  = 
\left( \frac{\sigma_0}{\rho g} \right)^2 \,.
\ee
This energy density is positive-definite. In the standard 
cosmological description, the scale factor is $a(t)=a_0 \sqrt{t-t_0}$ 
with an increasing function $a(t)$; for the ice cap model of Nye, the 
downward slope of the ice corresponds to $h'(x)<0$ and $0\leq x \leq L$. 
This profile is reflected about the $x=0$ axis to produce an overall 
profile symmetric under the reflection $x \to -x$ and with a cusp at 
$x=0$, where the left and right derivatives have opposite signs 
\cite{Paterson,CuffeyPaterson,Hooke,GreveBlatter}. The ice cap model corresponding to 
the correct Glen law~(\ref{Glen}) for ice rheology satisfies instead the 
Vialov 
equation \cite{Vialov,Paterson,CuffeyPaterson,Hooke,GreveBlatter}
\be
x \, c(x) = \frac{ 2 {\cal A}}{n+2} \left( \rho g h \left| \frac{dh}{dx} 
\right| \right)^n h^2 \label{Vialoveq}
\ee
where $c(x)$ describes the accumulation rate of ice per unit of area 
normal to  the vertical direction and per unit time (volume of ice added 
per unit time and per square meter, {\em i.e.}, a flux density), $n=3$ 
for ice creep, and ${\cal A}$ is the constant appearing in the Glen 
law~({\ref{Glen}). The 
Vialov profile is obtained for $c=$~const.$\equiv c_0$,
\be
h(x) = H \left[ 1-\left( \frac{x}{L} \right)^{ \frac{n+1}{n} } \right]^{ 
\frac{n}{2(n+1)} } \,,\label{Vialovprofile}
\ee
where $H=h(0)$, $h(L)=0$, and 
\be
L= \frac{H^2}{ 2^{ \frac{n}{n+1} } }\left[ \frac{2{\cal A}}{(n+2) c_0 } \, \left( 
\rho g \right)^n \right]^{\frac{1}{n+1} } 
\ee
(the Lagrangian formulation and cosmic analogy for this equation are 
presented in \cite{Faraoni:2020ips,mybook}). 

In the limit $n\to +\infty$ of plastic ice, the Vialov 
equation~(\ref{Vialoveq}) reduces to~(\ref{questa}) while the Vialov 
profile~(\ref{Vialovprofile}) becomes the parabolic Nye 
profile~(\ref{Nyeprofile})  \cite{Vialov,Paterson}.

\section{Lava dome on a uniform slope}
\label{sec:6}

 Consider now the flow of a Bingham fluid over a plane of uniform slope 
$\beta$, in stationary state and in the shallow lava approximation, 
building a dome on this slope \cite{OsmondGriffiths98}.  The same problem 
has been approached in glaciology by studying an ice sheet made of plastic 
ice on a slope \cite{EJP}, although for purely pedagogical purposes since 
the correct rheology is given by Glen's law~(\ref{Glen}) instead of 
plastic ice. The result of \cite{EJP} is 
\be
hh'-h\sin\beta + h_0=0 \, , \quad \quad h_0= \frac{ \tau_b}{\rho g} 
\,.\label{myequation}
\ee
This equation has the analytical solution (for arbitrary large slope 
angles  
$\beta$) \cite{EJP}
\be
x(h) =  L+ \frac{h}{\sin\beta} + \frac{h_0}{\sin^2 \beta} \, \ln \left( 
1-\frac{h}{h_0} \, \sin\beta \right) \,, \label{mysolution}
\ee
which satisfies the boundary condition $ h=0$ at $x=L$. 

There is a difference between ice caps and lava flows.  While 
precipitation on a glacier is distributed (as described by the function 
$c(x)$), lava erupts from a vent and one must describe both down-slope and 
up-slope flows from this vent, as is done in theoretical descriptions, 
which commonly leads to {\em two} solutions of the relevant differential 
equation \cite{Griffithsreview}. Although not considered in \cite{EJP}, 
the up-slope solution can be recovered by changing the sign of the basal 
stress $\tau_b$, therefore of $h_0$, in Eq.~(\ref{myequation}).  Osmond \& 
Griffiths  \cite{OsmondGriffiths98} 
study a silicic lava dome on a slope (silicic lava has relatively low 
temperature and high viscosity). The lava thickness $h(t, 
x, y )$ satisfies 
the equation \cite{OsmondGriffiths98,Griffithsreview}
\be
\left( \frac{\partial h_1}{\partial x} - \sin\beta \right)^2 
+\left( 
\frac{\partial h_1}{\partial y} \right)^2 = \left( \frac{ \sigma_0 }{\rho 
g h \cos\beta} \right)^2
\ee
where one assumes symmetry about the $y=0$ line which, by continuity, results in $ 
\partial h_1/\partial y=0 $ on the line $x=0$.  Here $h_1(x)$ is the 
vertical position of the lava, not its thickness $h$ 
\cite{OsmondGriffiths98,Griffithsreview}, to which it is related by $ h= 
h_1 \cos\beta $. Solving for the longitudinal lava profile along the 
$x=0$ line and using the rescaled variables 
\begin{eqnarray}
\bar{x} & \equiv & \left( \frac{ \rho g}{\sigma} \,  \sin^2\beta \cos\beta 
\right) x 
\,,\\
&&\nonumber\\
\bar{y} & \equiv & \left( \frac{ \rho g}{\sigma} \,  \sin^2\beta \cos\beta 
\right) y 
\,,\\
&&\nonumber\\
\bar{h}_1 &\equiv & \left( \frac{ \rho g}{\sigma} \,  \sin\beta \cos\beta 
\right) h 
\,,
\end{eqnarray}
Osmond \& Griffiths find the solution 
\cite{OsmondGriffiths98,Griffithsreview}
\be
x(h) = \left\{ \begin{array}{ll} 
\bar{h}_1-\bar{H}_1 +\ln \left( \frac{1-\bar{h}_1}{1-\bar{H}_1} \right) 
& \quad \mbox{if } \;\; x\geq 0 \,,\\
 &\\
\bar{h}_1-\bar{H}_1 -\ln \left( \frac{1+\bar{h}_1}{1+\bar{H}_1} \right) 
& \quad \mbox{if } \;\; x\leq 0 \,,
\end{array} \right. \label{OGsolution}
\ee
which satisfies the boundary condition $\bar{h}_1=\bar{H}_1$ at $x=0$, and 
where the flow has length $ \bar{L}_1= -\ln \left| 1-\bar{H}_1^2 \right| $ 
and width $\bar{W}_1 \simeq 2\left( 1-\sqrt{ 1-\bar{H}_1^2} \right)$ 
\cite{OsmondGriffiths98,Griffithsreview}.   
This solution coincides with the solution~(\ref{mysolution}) of \cite{EJP} 
for perfectly plastic ice and was rediscovered in Ref.~\cite{LiuMei89} 
together with the Nye profile (\ref{Nyeprofile}).

The cosmic analogy is obtained by rewriting Eq.~(\ref{mysolution})  as
\be
\left( \frac{h'}{h} \right)^2 = \frac{\sin^2 \beta}{h^2} + 
\frac{h_0^2}{h^4} -\frac{2h_0 \sin\beta}{h^3}  \,,
\ee
which is analogous to the Friedmann equation
\be
H^2 = -\frac{K}{a^2} + \frac{8\pi G}{3} \left( 
\frac{ \rho^{(0)}_\mathrm{(rad)} }{a^4}  
+ \frac{ \rho^{(0)}_\mathrm{(dust)} }{a^3} \right)
\ee
where 
\begin{eqnarray}
K &=& -\sin^2 \beta \,,\\
&&\nonumber\\
\frac{8\pi G }{3} \, \rho^{(0)}_\mathrm{(rad)} &=& h_0^2 \,,\\
&&\nonumber\\
\frac{8\pi G }{3} \, \rho^{(0)}_\mathrm{(dust)} &=& -2 h_0\sin\beta <0  
\,.
\end{eqnarray}
The effective Lagrangian and Hamiltonian for the lava flow obtained from 
the analogy are
\begin{eqnarray}
L &=& h \, h'^2 + h \sin^2 \beta +\frac{h_0^2}{h} -2h_0 \sin\beta  \,,\\
&&\nonumber\\
{\cal H} &=& h \, h'^2 - h \sin^2 \beta - \frac{h_0^2}{h} +2h_0 \sin\beta  
\,,
\end{eqnarray}
and Eq.~(\ref{myequation}) is obtained by imposing the Hamiltonian 
constraint of general relativity ${\cal H}=0$. 

The width of the lava flow in the transverse $y$-direction is obtained 
\cite{OsmondGriffiths98} for $\partial \bar{h}_1 /\partial \bar{x}\simeq 0$, 
which leads to 
\be
1+ \left( \frac{ \partial \bar{h}_1}{ \partial \bar{y} } \right)^2 = 
\frac{1}{ \bar{h}_1^2} 
\ee
and to another cosmic analogy through the analogue of the Friedmann equation
\be
\left( \frac{ \bar{h}_1 ' }{ \bar{h}_1} \right)^2 = -\frac{1}{ \bar{h}_1^2} 
+\frac{1}{\bar{h}_1^4}  
\ee
(where now a prime denotes differentiation with respect to $\bar{y}$), which describes a spatially closed ($K=+1$) universe sourced by blackbody radiation. The solution of \cite{OsmondGriffiths98}
\be
\bar{y} ( \bar{h}_1 )= \pm \left( \sqrt{ 1-\bar{h}_1^2} - 
\sqrt{ 1-\bar{H}_1^2} \, \right) \,,
\ee
which can be inverted as
\be
\bar{h}_1( \bar{y} ) = \sqrt{ 1- \left( \bar{y} \mp \sqrt{ 1- \bar{H}_1^2 }\, 
\right)^2 } \,,
\ee
is a classic solution of FLRW cosmology \cite{Faraoni:2021opj} and can be rewritten
 in the parametric form
\begin{eqnarray}
\bar{h}_1(\eta) &=& \sin\eta \,,\\
&&\nonumber\\
\bar{y}(\eta) &=& \pm \sqrt{ 1-\bar{H}_1^2} + \cos\eta \,,
\end{eqnarray}
where the parameter $\eta $ corresponds to the conformal time of FLRW cosmology defined 
by $dt=a  d\eta$. This FLRW universe begins at a Big Crunch, reaches a maximum size, and then shrinks and collapses to a Big Crunch, mirroring the transverse profile of the lava dome of finite extension. The corresponding Lagrangian and Hamiltonian are 
\begin{eqnarray}
L_1 \left( \bar{h}, \bar{h}' \right) &=& \bar{h}  \, \bar{h}'^2
-\bar{h} +\frac{1}{ \bar{h} } \,,\\
&&\nonumber\\
{\cal H}_1 \left(\bar{h}, \bar{h}' \right) &=& \bar{h} \, \bar{h}'^2
+\bar{h} - \frac{1}{ \bar{h} } \,.
\end{eqnarray}

\section{Conclusions}
\label{sec:7}

The Friedmann equation~(\ref{eq:11}) lends itself to various analogies 
\cite{mybook}, including the differential equations ruling lava flows, 
because it resembles the energy conservation equation for a 
one-dimensional motion. Analogue gravity, in which laboratory scale 
systems mimic gravitational systems such as black holes, wormholes, and 
universes that cannot be recreated in the lab, has become a mature area of 
science ({\em e.g.}, 
\cite{Barcelo:2005fc,Volovik:2003fe,Belgiorno:2019ofm,Visser:2001fe,Liberati:2017jnr}). 
Analogue gravity 
systems comprise Bose-Einstein condensates and other condensed matter 
systems 
\cite{Barcelo:2000tg,Fedichev:2003id,Barcelo:2003et,Fedichev:2003bv,Fischer:2004bf,Cha:2016esj,Eckel:2017uqx,Fedichev:2003dj,Volovik:1997mtb,Jacobson:1998he,Volovik:1999mj,CM4,Volovik:2002ci,Pashaev:1998sk}, fluids 
\cite{Unruh:1980cg,Unruh:1994je,Visser:1997ux,Garay:2000jj,Fischer:2001jz,fluid6,Nandi:2004nw,Visser:2004zs,Slatyer:2005ty,Weinfurtner:2010nu,Torres:2016iee,Patrick:2018orp,Patrick:2019kis}, 
optical systems 
\cite{Schutzhold:2001fw,Unruh:2003ss,Davis:2003qfo,Schutzhold:2004tv,Prain:2019jqk} 
and even soap bubbles \cite{CriadoAlamo,bubble2} and capillary flow 
\cite{capillary}. Specifically, analogues of FLRW cosmology have been 
found in Bose-Einstein condensates 
\cite{acosmo1,Volovik:2000ua,Prain:2010zq,Braden:2019vsw}. Based on the 
cosmic analogies presented here, 
tabletop experiments studying the flow of Bingham fluids of interest in 
geophysics, which employ slurries descending inclines \cite{Griffithsreview}, 
can constitute analogue gravity systems 
for cosmology. The most interesting phenomena discovered in analogue 
gravity thus far involve wave propagation and perturbations, which have 
led to the discovery of analogue Hawking radiation \cite{Weinfurtner:2010nu}, 
superradiance \cite{Slatyer:2005ty}, and cosmological particle production 
\cite{Fedichev:2003bv}, predicted in quantum field theory in curved 
spacetime and not directly observable in nature, but other aspects may 
be disclosed by analogue gravity in the future.

As seen above, negative energy densities for the cosmological fluids 
analogous to geophysical flows do occur sometimes and they are responsible 
for the appearance of hyperbolic functions in the scale factor $a(t)$ 
(cosmologists are familiar with hyperbolic functions in the presence of 
a negative cosmological constant or of positive curvature index $K$ 
\cite{MTW,Wald,Carroll,Liddle,KT}). While these negative 
densities would be unacceptable for real fluids in Einstein gravity, they 
can be viable 
in alternative theories where extra degrees of freedom with respect to 
general relativity can be described as {\em effective} fluids not subject 
to the usual physical requirements imposed on ordinary fluids ({\em e.g.}, 
\cite{Faraoni:2018qdr,Giusti:2021sku,Miranda:2022wkz}). The extension of 
the cosmological analogies reported here to scalar-tensor and other 
theories of gravity alternative to general relativity will be pursued 
elsewhere.

\begin{acknowledgements}

This work is supported by Bishop's University and by the Natural Sciences 
\& Engineering Research Council of Canada (grant 2016-03803).

\end{acknowledgements}

%\bigskip
%\appendix
%\section{TITLE OF THE APPENDIX}
%\label{AppendixA}
%\renewcommand{\theequation}{A.\arabic{equation}}



\end{document}